\documentstyle[amssymb,epsf,amsmath,amsfonts,twocolumn,prl,aps]{revtex}

\tighten

\begin{document}
\draft

\twocolumn[\hsize\textwidth\columnwidth\hsize\csname
@twocolumnfalse\endcsname

\title{Slowly rotating charged fluid balls and their matching to an exterior domain}
\author{Gyula Fodor$^{1}$, Zolt{\'{a}}n Perj{\'{e}}s$^{1}$ and Michael Bradley$^{2}$} 
 \address{$1$ KFKI Research Institute for Particle and Nuclear Physics,\\H--1525, Budapest 114, P.O.B.\ 49, Hungary\\
$2$ Department of Plasma Physics, Ume{\aa } University, S--901 87, Ume{\aa },
Sweden }
 \date{\today}
\maketitle

\begin{abstract}
The slow-rotation approximation of Hartle is developed to a setting where a
charged rotating fluid is present. The linearized Einstein-Maxwell equations
are solved on the background of the Reissner-Nordstr\"{o}m space-time in the
exterior electrovacuum region. The theory is put to action for the charged
generalization of the Wahlquist solution found by Garc\'{\i}a. The Garc\'{\i}a
solution is transformed to coordinates suitable for the matching and expanded
in powers of the angular velocity. The two domains are then matched along the
zero pressure surface using the Darmois-Israel procedure. We prove a theorem
to the effect that the exterior region is asymptotically flat if and only if
the parameter $C_{2}$, characterizing the magnitude of an external magnetic
field, vanishes. We obtain the form of the constant $C_{2}$ for the
Garc\'{\i}a solution. We conjecture that the Garc\'{\i}a metric cannot be
matched to an asymptotically flat exterior electrovacuum region even to first
order in the angular velocity. This conjecture is supported by a high
precision numerical analysis.
\end{abstract}

\pacs{PACS:numbers: 04.40.Nr, 04.25.-g, 04.20.Cv, 04.20.Jb, 04.40.Dg}

\vspace{2ex}
]

\section{\smallskip Introduction}\label{introduction}

There are a surprizingly few rotating perfect fluid solutions of Einstein's
relativistic equations known to date. One of the most comprehensive of these
is the rigidly rotating charged perfect fluid solution of
Garc\'{\i}a\cite{garc}. This metric is type D in the Petrov classification of
the curvature. The Einstein-Maxwell equations are satisfied with a stress
tensor which is the sum of that of the perfect fluid and the Maxwell field.
The fluid medium carries electric charge, and it has a divergence-free four-current.

Because of the tantalizing lack of explicit, relativistically rotating fluid
stellar models in the literature, it is only natural to ask if it is possible
to join a suitably chosen domain of the Garc\'{\i}a solution to an external
source-free Einstein-Maxwell region.

The Garc\'{\i}a solution reduces to the Wahlquist space-time in the absence of
electric charge. Unfortunately, as we have proven in an earlier
publication\cite{bfmp}, the Wahlquist solution is unsuitable as a model of an
isolated relativistic object. \textit{A priori} it is unclear if the presence
of electric charge can supply the necessary ingredients for a smooth matching.

The purpose of the present paper is to search for the conditions of matching
the Garc\'{\i}a metric to an asymptotically well-behaved external
electrovacuum domain. A suitable vehicle for performing this research is the
slow-rotation approximation. We could successfully use this approach in our
previous computations for the Wahlquist space-time\cite{bfmp}. The foundations
of the slow-rotation approximation were laid down by Hartle\cite{Hartle}. In
its original form, the Hartle method matches an internal fluid domain with an
external vacuum. An essential tenet of this approach is the independence of
the first-order perturbation function $\omega$ of the polar angle $\vartheta.$
For a generic setting in the presence of the Maxwell field, one would have to
go through the proof of this independence. In the present case, however, we
can omit this general theory since $\omega$ is a function of the radius alone
in the slowly rotating Garc\'{\i}a space-time. As a consequence of the
matching conditions, then, this property is inherited by the electrovacuum exterior.

In this paper, we apply the slow-rotation approximation to the Garc\'{\i}a
space-time. Our result is that the electric charge works against the
conditions of matching. We find that matching to an asymptotically flat
electrovacuum exterior is impossible already to the first order in the angular
velocity. This comes as a surprize to us since all uncharged fluids can be
matched to an asymptotically flat vacuum exterior to first order in the
rotation parameter. This is because such vacuum space-times do not differ from
the first-order form of the the Kerr metric.

The paper is organized as follows. In the next section, we compute the
Garc\'{\i}a metric in various forms that will be necessary to launch the
matching process. In subsection \ref{static} we take the static limit which, of course,
is spherically symmetric. Next, in subsection \ref{combined} we assemble the slowly
rotating form of the metric. In Sec. \ref{electrovacuum} we prepare the field quantities in the
external electrovacuum domain. Thus we get the general solution for the
first-order perturbation function $\omega,$ and we investigate its asymptotic
properties. The actual matching of the two domains takes place in Sec.  \ref{match}. As
before, we require that the Darmois-Israel conditions\cite{israel} are
satisfied. The details are worked out in subsection \ref{cond}, generalizing the
theory in the presence of a Maxwell field. The solution of the matching
conditions for the static state and to first order in the angular velocity is
carried out in subsections \ref{sphere} and \ref{first}, respectively. The results are further
discussed in Sec. \ref{discussion}. The Appendix is devoted to the issue of the choice of
independent parameters.

\section{The interior metric}\label{interior}

The charged generalization of the Wahlquist solution given by Garc\'{\i}a
\cite{garc} has the metric
\begin{align}
ds^{2}&=\frac{P}{\Delta}\left(  d\tau+\delta Nd\sigma\right)  ^{2}-\frac
{Q}{\Delta}\left(  d\tau+\delta Md\sigma\right)  ^{2}\nonumber\\
&+\Delta\left(\frac{dx^{2}}{P}+\frac{dy^{2}}{Q}\right) \label{dsgarc}%
\end{align}
with the real functions
\begin{align}
\Delta & =M-N\nonumber\\
M  & =\frac{1}{k^{2}}\sinh^{2}(kx)-\xi_{0}^{2}\nonumber\\
N  & =-\frac{1}{k^{2}}\sin^{2}(ky)-\xi_{0}^{2}\label{fcs}\\
P  & =a+\tfrac{1}{2k}\left[  2n+x\left(  \alpha+\beta^{2}\right)  \right]
\sinh(2kx)\nonumber\\
&+\left[  b-\left(  g+\beta x\right)  ^{2}\right]  \cosh
(2kx)\nonumber\\
Q  & =a-\tfrac{1}{2k}\left[  2m+y\left(  \alpha+\beta^{2}\right)  \right]
\sin(2ky)\nonumber\\
&+\left[  b+\left(  e+\beta y\right)  ^{2}\right]  \cos(2ky)\ .\nonumber
\end{align}
Here $a,$ $b,$ $e,$ $g,$ $k,$ $m,$ $n,$ $\alpha,$ $\beta,$ $\delta$ and
$\xi_{0}$ are eleven constants. The number of independent parameters modulo
diffeomorphisms is eight (\textit{Cf.} Appendix). This metric is a solution of
Einstein's equations $G_{\alpha\beta}=8\pi\left(  T_{\alpha\beta}%
^{(f)}+T_{\alpha\beta}^{(e)}\right)  $ with
\begin{align*}
T_{\alpha\beta}^{(f)}  & =(\mu+p)u_{\alpha}u_{\beta}+pg_{\alpha\beta}\\
T_{\alpha\beta}^{(e)}  & =\frac{1}{4\pi}\left(  F_{\alpha\gamma}F_{\beta
}^{\ \gamma}-\frac{1}{4}g_{\alpha\beta}F_{\gamma\delta}F^{\gamma\delta
}\right)
\end{align*}
where $p$ is the pressure, $\mu$ the density, $u^{\alpha}$ the velocity of the
fluid, and $F_{\alpha\beta}$ is the electromagnetic field tensor. In comoving
coordinates, the velocity vector is $u=u^{0}\partial/\partial\tau$. The
pressure and density are given by
\begin{align*}
8\pi p  & =-\frac{k^{2}}{\Delta}\left(  Q-P\right)  +\alpha k^{2}+\Sigma\\
8\pi\mu & =3\frac{k^{2}}{\Delta}\left(  Q-P\right)  -\alpha k^{2}-\Sigma\\
\Sigma & =-\frac{2\beta k}{\Delta}\left[  \left(  e+\beta y\right)
\sin(2ky)+\left(  g+\beta x\right)  \sinh(2kx)\right]  \ .
\end{align*}
The electromagnetic four-potential $A_{\alpha}$ determines the field
tensor\footnote{Garc\'{\i}a\cite{garc} uses the definition $F_{\alpha\beta
}=2A_{[\alpha,\beta]},$ which gives rise to an overall sign difference in
$A_{\alpha}.$} as $F_{\alpha\beta}=\partial_{\alpha}A_{\beta}-\partial_{\beta
}A_{\alpha},$ and its components are
\begin{align*}
A_{x}  & =A_{y}=0\\
A_{\sigma}  & =-\frac{\delta}{2\Delta k}\left[  \left(  e+\beta y\right)
M\sin(2ky)+\left(  g+\beta x\right)  N\sinh(2kx)\right] \\
A_{\tau}  & =-\frac{1}{2\Delta k}\left[  \left(  e+\beta y\right)
\sin(2ky)+\left(  g+\beta x\right)  \sinh(2kx)\right]\\ 
& =\frac{\Sigma}%
{4k^{2}\beta}\ .
\end{align*}
The electric current vector $j^{\alpha}=\rho u^{\alpha}$ satisfies the field
equation $4\pi j^{\alpha}=F_{\ \ \ ;\beta}^{\alpha\beta}$. The current $4\pi
j^{\alpha}=2k^{2}\beta\delta_{\tau}^{\alpha}$ has no divergence (as it should,
since the Maxwell equations are satisfied) and the conserved charge density
is
\[
\rho=\frac{k^{2}\beta}{2\pi}\sqrt{\frac{Q-P}{\Delta}\ .}%
\]
Note that this solution has been named the \emph{Wahlquist-Newman metric} in
\cite{mars}.

\subsection{The Wahlquist form}\label{ Wahlquist}

Our goal in this section is to bring the metric (\ref{dsgarc}) to a form which
reduces to that of the Wahlquist solution in the no-charge limit. The static
limit of the Wahlquist form will, in turn, yield the charged generalization of
the Whittaker space-time. To achieve this, we transform to the new coordinates
$\xi$ and $\zeta$ by setting
\begin{equation}
k\xi=\sinh(kx)\ \ ,\ \ \ k\zeta=\sin(ky).\label{xy2xizeta}%
\end{equation}
Then we have
\begin{equation}
\Delta=\xi^{2}+\zeta^{2}\ .
\end{equation}
In order to be able to introduce Wahlquist's functions $h_{1}$ and $h_{2},$ we
rescale the functions $P$ and $Q$ using a new constant parameter $r_{0};$%
\[
h_{1}=r_{0}^{2}Q\ \ ,\ \ \ h_{2}=r_{0}^{2}P.
\]
\newline To choose the factors of the other terms appropriately, we
also rescale the $\tau$ and $\sigma$ coordinates as
\[
{\varphi}=-\frac{1}{r_{0}^{2}}\sigma\ \ ,\ \ \ t=\frac{1}{r_{0}}\tau\ .
\]
The metric takes the form
\begin{align}
ds^{2}  & =-\frac{h_{1}-h_{2}}{\Delta}\left(  dt-{\mathbf{A}}d{\varphi}\right)
^{2}+r_{0}^{2}\delta^{2}\Delta\frac{h_{1}h_{2}}{h_{1}-h_{2}}d{\varphi}%
^{2}\nonumber\\
& +\Delta r_{0}^{2}\left[  \frac{d\zeta^{2}}{\left(  1-k^{2}\zeta^{2}\right)
h_{1}}+\frac{d\xi^{2}}{\left(  1+k^{2}\xi^{2}\right)  h_{2}}\right]
\label{dswahl}%
\end{align}
where $\mathbf{A}$ is Wahlquist's function,
\begin{equation}
{\mathbf{A}}=\delta r_{0}\left(  \frac{h_{1}\xi^{2}+h_{2}\zeta^{2}}{h_{1}-h_{2}%
}-\xi_{0}^{2}\right)  \ .\label{calA}%
\end{equation}
We introduce the rescaled constants
\begin{align}
\bar{m}  & =mr_{0}^{3}\ \ ,\ \ \ \bar{b}=-nr_{0}^{3}\\
\bar{e}  & =er_{0}\ \ ,\ \ \ \bar{g}=gr_{0}\ \ ,\ \ \ \bar{\beta}=\beta
r_{0}\nonumber
\end{align}
and three new constants as follows. In place of $\alpha$ we shall use the
constant $\kappa$, the constant $a$ will be replaced by $C$ and finally $E$
will take the place of the constant $b,$ using the following
definitions\footnote{We have slight differences from the notation of
Garc\'{\i}a\cite{garc}: $\overline{m}$ and $\overline{b}$ have different
factors, $\kappa$ includes a term $\beta^{2}$, and $E$ and $C$ are not set to
$1$ yet}
\begin{align}
\frac{1}{\kappa^{2}}  & =r_{0}^{2}\left(  \alpha+\beta^{2}\right) \nonumber\\
C  & =r_{0}^{2}(a+b)\\
E  & =-2r_{0}^{2}bk^{2}-\frac{1}{\kappa^{2}}.\nonumber
\end{align}
We get for the functions
\begin{align}
h_{1}  & =\frac{\zeta}{\kappa^{2}}\left[\zeta-\frac{1}{k}\sqrt{1-k^{2}\zeta^{2}}
\arcsin(k\zeta)\right]-\frac{2\bar{m}}{r_{0}}\zeta\sqrt{1-k^{2}\zeta^{2}} \nonumber\\
& +\left[  \bar{e}+\frac{\bar{\beta}}{k}\arcsin(k\zeta)\right]^{2}%
\left(1-2k^{2}\zeta^{2}\right)+C+E\zeta^{2}\\
h_{2}  & =-\frac{\xi}{\kappa^{2}}\left[  \xi-\frac{1}{k}\sqrt{1+k^{2}\xi^{2}}\mathrm{arcsinh}%
(k\xi)\right]-\frac{2\bar{b}}{r_{0}}\xi\sqrt{1+k^{2}\xi^{2}} \nonumber\\
& -\left[  \bar{g}+\frac{\bar{\beta}}{k}\mathrm{arcsinh}(k\xi)\right]^{2}%
\left(  1+2k^{2}\xi^{2}\right)+C-E\xi^{2}  \ .
\end{align}

The electromagnetic potential 1-form $A=A_{\alpha}dx^{\alpha}$ is
\begin{align}
A  & =-\frac{1}{\Delta}\left[  \bar{g}+\frac{\bar{\beta}}{k}\mathrm{arcsinh}%
(k\xi)\right]  \xi\sqrt{1+k^{2}\xi^{2}} \nonumber\\
&\times\left[  dt+\left(  \zeta^{2}+\xi
_{0}^{2}\right)  \delta r_{0}d{\varphi}\right] \nonumber\\
& -\frac{1}{\Delta}\left[  \bar{e}+\frac{\bar{\beta}}{k}\arcsin(k\zeta
)\right]  \zeta\sqrt{1-k^{2}\zeta^{2}} \nonumber\\
&\times\left[  dt-\left(  \xi^{2}-\xi_{0}%
^{2}\right)  \delta r_{0}d{\varphi}\right]  .
\end{align}
The pressure and density become
\begin{align}
8\pi p  & =-\frac{k^{2}}{\Delta r_{0}^{2}}\left(  h_{1}-h_{2}\right)
+\frac{k^{2}}{r_{0}^{2}\kappa^{2}}\left(  1-\bar{\beta}^{2}\kappa^{2}\right)
+\Sigma\label{pWahl}\\
8\pi\mu & =3\frac{k^{2}}{\Delta r_{0}^{2}}\left(  h_{1}-h_{2}\right)
-\frac{k^{2}}{r_{0}^{2}\kappa^{2}}\left(  1-\bar{\beta}^{2}\kappa^{2}\right)
-\Sigma\label{muWahl}\\
\Sigma & =\frac{4\bar{\beta}k^{2}}{r_{0}^{2}}A_{t}\ .\label{mup}%
\end{align}

The expressions (\ref{dswahl})-(\ref{mup}) contain a large number of
parameters: $C,$ $E,$ $\bar{m},$ $\bar{b},$ $k,$ $\kappa,$ $\bar{e},$ $\bar
{g},$ $\bar{\beta},$ $r_{0},$ $\delta$ and $\xi_{0}.$ How many of these twelve
constants are necessary to uniquely describe the metric? This question will
next be addressed.

By using linear coordinate transformations of $t$ and ${\varphi}$ we can
obviously set $\delta=1$ and $\xi_{0}=0.$ The parameter $r_{0}$ has been
introduced in order to enable ourselves to go to the slow-rotation limit. The
scaling of this parameter can be chosen arbitrarily. Defining $r_{0}^{\prime
}=cr_{0},$ $C^{\prime}=c^{2}C,$ $E^{\prime}=c^{2}E,$ $\bar{m}^{\prime}%
=c^{3}\bar{m},$ $\bar{b}^{\prime}=c^{3}\bar{b},$ $\kappa^{\prime}=\kappa/c,$
$\bar{e}^{\prime}=c\bar{e},$ $\bar{g}^{\prime}=c\bar{g},$ $\bar{\beta}%
^{\prime}=c\bar{\beta},$ where $c$ is a constant, the $d\zeta^{2}$ and
$d\xi^{2}$ terms in the metric remain unchanged with the primed quantities.
However, we need to perform a transformation of the coordinates $t=ct^{\prime
},$ ${\varphi}=c^{2}{\varphi}^{\prime}$ to keep the structure of the other
components. This shows that the two sets of constants $\left\{  C,E,\bar
{m},\bar{b},k,\kappa,\bar{e},\bar{g},\bar{\beta},r_{0}\right\}  $ and
$\left\{  C^{\prime},E^{\prime},\bar{m}^{\prime},\bar{b}^{\prime}%
,k,\kappa^{\prime},\bar{e}^{\prime},\bar{g}^{\prime},\bar{\beta}^{\prime
},r_{0}^{\prime}\right\}  $ provide parametrizations of a single physical
state equivalent under the diffeomorphism $\left\{  \zeta^{\prime},\xi
^{\prime},{\varphi}^{\prime},t^{\prime}\right\}  =\left\{  \zeta,\xi
,c^{-2}{\varphi},c^{-1}t\right\}  $.

Another freedom of choice follows from the coordinate transformation
$\zeta^{^{\prime\prime}}=f\zeta,$ $\xi^{^{\prime\prime}}=f\xi,$ where $f$ is a
constant. In the metric (\ref{dswahl}) we have
\begin{equation}
\Delta=\xi^{2}+\zeta^{2}=\frac{1}{f^{2}}\left(  \xi^{^{\prime\prime}2}%
+\zeta^{^{\prime\prime}2}\right)  =\frac{1}{f^{2}}\Delta^{^{\prime\prime}}\ .
\end{equation}
Introducing the constants $k^{^{\prime\prime}}=k/f,$ $C^{^{\prime\prime}%
}=f^{4}C,$ $E^{^{\prime\prime}}=f^{2}E,$ $\bar{m}^{^{\prime\prime}}=f^{3}%
\bar{m},$ $\bar{b}^{^{\prime\prime}}=f^{3}\bar{b},$ $\kappa^{^{\prime\prime}%
}=\kappa/f,$ $\bar{e}^{^{\prime\prime}}=f^{2}\bar{e},$ $\bar{g}^{^{\prime
\prime}}=f^{2}\bar{g},$ $\bar{\beta}^{^{\prime\prime}}=f\bar{\beta},$ and
defining $h_{1}^{^{\prime\prime}}$ and $h_{2}^{^{\prime\prime}}$ similarly to
$h_{1}$ and $h_{2},$ we get
\begin{equation}
h_{1}=\frac{1}{f^{4}}h_{1}^{^{\prime\prime}}\ \ ,\ \ \ h_{2}=\frac{1}{f^{4}%
}h_{2}^{^{\prime\prime}}\ .
\end{equation}
Then the $d\zeta^{2}$ and $d\xi^{2}$ terms transform appropriately. However,
to get the correct transformation for the other terms in the metric, we need
to change the coordinates as $t=ft^{^{\prime\prime}}$ and ${\varphi}%
=f^{3}{\varphi}^{^{\prime\prime}}.$ It follows that the two sets of constants,
$\left\{  C,E,\bar{m},\bar{b},k,\kappa,\bar{e},\bar{g},\bar{\beta}%
,r_{0}\right\}  $ and $\left\{  C^{^{\prime\prime}},E^{^{\prime\prime}}%
,\bar{m}^{^{\prime\prime}},\bar{b}^{^{\prime\prime}},k^{^{\prime\prime}%
},\kappa^{^{\prime\prime}},\bar{e}^{^{\prime\prime}},\bar{g}^{^{\prime\prime}%
},\bar{\beta}^{^{\prime\prime}},r_{0}\right\}  $ are equivalent descriptors.

At this stage, the coordinates are completely fixed. Hence there are
\emph{eight} independent physical parameters. The corresponding analysis of
the parameters in the original coordinates is not needed in the present paper,
but for convenience, is included as an Appendix. Using the coordinate
transformations $x^{i\prime}$ = $x^{i\prime}\left(  x^{k}\right)  $ and
$x^{i\prime\prime}$ = $x^{i\prime\prime}\left(  x^{k}\right)  $, for example,
the constants $C$ and $E$ can always be made $\pm1$ or $0.$ Garc\'{\i}a
\cite{garc} discusses the case $C=E=1$, which for zero $\bar{e},$ $\bar{g}$
and $\bar{\beta}$ yields the original uncharged Wahlquist solution. When $C$
or $E$ can only be rescaled to $-1$ or $0,$ some different solutions might
arise, as was first noted by Mars and Senovilla \cite{mase} in the uncharged case.

There exist two combinations of these transformations which preserve many of
the constants. Table 1. shows the sets of constants giving metrics equivalent
to the original set $\left\{  C,E,\bar{m},\bar{b},k,\kappa,\bar{e},\bar
{g},\bar{\beta},r_{0}\right\}  ,$ when choosing $cf=1$ and $cf^{2}=1$ .%

\begin{align*}
& \text{\textit{TABLE I}. Equivalent sets of parameters }\\
&
\begin{tabular}
[c]{c|cccccccccc}\hline\hline
$1$ & $C$ & $E$ & $\bar{m}$ & $\overset{}{\bar{b}_{{}}^{{}}}$ & $k$ & $\kappa$
& $\bar{e}$ & $\underset{}{\bar{g}}$ & $\bar{\beta}^{{}}$ & $r_{0}^{{}}%
$\\\hline\hline
$cf=1$ & $c^{2}C$ & $E$ & $\bar{m}$ & $\bar{b}^{{}}$ & $\frac{1^{{}}}{c_{{}}%
}k$ & $\kappa$ & $c\bar{e}$ & $c\bar{g}$ & $\overset{}{\bar{\beta}}$ &
$\frac{1}{\underset{}{c}}r_{0}^{{}}$\\\hline
$cf^{2}=1$ & $C$ & $f^{2}E$ & $f^{3}\bar{m}$ & $f^{3}\bar{b}$ & $fk$ &
$\frac{1^{{}}}{\underset{}{f_{{}}}}\kappa$ & $\bar{e}$ & $\bar{g}$ &
$\overset{}{f\bar{\beta}}$ & $f^{2}r_{0}$\\\hline\hline
\end{tabular}
\end{align*}

\textit{\ }

\subsection{The static charged fluid\label{sect}}\label{static}

In this section we consider the static limiting case of the Garc\'{\i}a
space-time. Following the procedure of Wahlquist\cite{wahl1}, we introduce a
new radial coordinate
\begin{equation}
r=\zeta r_{0}\label{zeta2r}%
\end{equation}
and a new constant $\gamma$ in place of $k$ by substituting
\begin{equation}
k=\gamma r_{0}\label{k2gamma}%
\end{equation}
everywhere. The constant $\gamma$ is related to Wahlquist's $\rho_{s}$ by
$\gamma^{2}=\kappa^{2}\rho_{s}$ . The static limit can be obtained then by
going to zero with $r_{0}.$ In the limit $r_{0}\rightarrow0$,
\begin{equation}
\lim_{r_{0}\rightarrow0}\left(  \frac{r_{0}^{2}}{r^{2}}h_{1}\right)
=\tilde{h}_{1}%
\end{equation}
where
\begin{align}
\tilde{h}_{1}  & =E-\frac{2\bar{m}}{r}\sqrt{1-\gamma^{2}r^{2}}\\
& +\frac{1}{\kappa^{2}}\left[  1-\frac{1}{\gamma r}\sqrt{1-\gamma^{2}r^{2}%
}\arcsin(\gamma r)\right] \nonumber\\
& +\left[  \frac{\tilde{e}}{r}+\frac{\bar{\beta}}{\gamma r}\arcsin(\gamma
r)\right]  ^{2}\left(  1-2\gamma^{2}r^{2}\right)  \ \nonumber
\end{align}
and $\tilde{e}=r_{0}\bar{e}.$ The function $h_{2}$ has the limiting form
\begin{equation}
\tilde{h}_{2}=\lim_{r_{0}\rightarrow0}h_{2}=C-E\xi^{2}-2\tilde{b}\xi-\left(
\bar{g}+\bar{\beta}\xi\right)  ^{2}\label{h2t}%
\end{equation}
where $\tilde{b}=\bar{b}/r_{0}.$ Using the limits $r_{0}^{2}\Delta\rightarrow
r^{2},$ and ${\mathbf{A}}\rightarrow0,$ the metric becomes
\begin{equation}
ds^{2}=-\tilde{h}_{1}dt^{2}+\frac{dr^{2}}{\left(  1-\gamma^{2}r^{2}\right)
\tilde{h}_{1}}+r^{2}\left(  \frac{d\xi^{2}}{\tilde{h}_{2}}+\delta^{2}\tilde
{h}_{2}d{\varphi}^{2}\right)  \ .
\end{equation}

A simpler form arises if we introduce a new radial coordinate $z$ by setting
\begin{equation}
\gamma r=\sin z.\label{r2z}%
\end{equation}
Then
\begin{align}
\tilde{h}_{1} &=E-2\bar{m}\gamma\cot z+\frac{1}{\kappa^{2}}\left(  1-z\cot
z\right)   \nonumber\\
& +\left(  \tilde{e}\gamma+\bar{\beta}z\right)  ^{2}\left(  \cot
^{2}z-1\right)  \ \label{h1tg}%
\end{align}
and
\begin{equation}
ds^{2}=-\tilde{h}_{1}dt^{2}+\frac{dz^{2}}{\gamma^{2}\tilde{h}_{1}}+\frac
{\sin^{2}z}{\gamma^{2}}\left(  \frac{d\xi^{2}}{\tilde{h}_{2}}+\delta^{2}%
\tilde{h}_{2}d{\varphi}^{2}\right)  \ .\label{statds}%
\end{equation}
The only nonvanishing component of the electromagnetic potential is
\[
A_{t}=-\left(  \tilde{e}\gamma+\bar{\beta}z\right)  \cot z\ .
\]
The pressure (\ref{pWahl}) and the density (\ref{muWahl}) are
\begin{align}
8\pi p  & =\gamma^{2}\left(  -\tilde{h}_{1}+\frac{1}{\kappa^{2}}-\bar{\beta
}^{2}\right)  +\Sigma\nonumber\\
8\pi\mu & =\gamma^{2}\left(  3\tilde{h}_{1}-\frac{1}{\kappa^{2}}+\bar{\beta
}^{2}\right)  -\Sigma\\
\Sigma & =4\bar{\beta}\gamma^{2}A_{t}\ .\nonumber
\end{align}

The bracketed term in (\ref{statds}) is the metric of a two-surface. We next
show that there is a parameter range for which this two-surface is the
two-sphere $S^{2}.$ In order to get the metric of the two-sphere, we introduce
a new coordinate $\theta$ in place of $\xi$ by $\delta^{2}\tilde{h}_{2}%
=c_{1}\sin^{2}\theta,$ where $c_{1}$ is some constant. Then
\begin{equation}
\delta^{2}\frac{d\tilde{h}_{2}}{d\xi}=2c_{1}\sin\theta\cos\theta\frac{d\theta
}{d\xi}%
\end{equation}
and
\begin{align}
\frac{d\xi^{2}}{\tilde{h}_{2}}&=\frac{4c_{1}^{2}\sin^{2}\theta\cos^{2}\theta
}{\delta^{4}\tilde{h}_{2}}\left(  \frac{d\tilde{h}_{2}}{d\xi}\right)
^{-2}d\theta^{2} \nonumber\\
&=\frac{4\left(  c_{1}-\delta^{2}\tilde{h}_{2}\right)  }%
{\delta^{2}}\left(  \frac{d\tilde{h}_{2}}{d\xi}\right)  ^{-2}d\theta^{2}\ .
\end{align}
The condition
\begin{equation}
\frac{d\xi^{2}}{\tilde{h}_{2}}=c_{2}d\theta^{2}%
\end{equation}
where $c_{2}$ is another constant, is satisfied if and only if
\begin{align}
c_{2}  & =\frac{1}{E+\bar{\beta}^{2}}\label{c22}\\
c_{1}  & =\delta^{2}\left[  C-\bar{g}^{2}+\frac{\left(  \tilde{b}+\bar{g}%
\bar{\beta}\right)  ^{2}}{E+\bar{\beta}^{2}}\right]  \ .\nonumber
\end{align}
We can use the map in the third row of Table 1 to set $c_{2}=1$ by making
$E+\bar{\beta}^{2}=1.$ Of course, this can be done only when $E+\bar{\beta
}^{2}$ is positive. The Lorentzian signature of the metric (\ref{statds}) hold
whenever it is possible to set
\begin{equation}
C-\bar{g}^{2}+\frac{\left(  \tilde{b}+\bar{g}\bar{\beta}\right)  ^{2}}%
{E+\bar{\beta}^{2}}=1
\end{equation}
using the second row of Table 1. Finally we set $\delta=1$ by rescaling the
${\varphi}$ coordinate. With these choices $\tilde{h}_{2}=1-\left(  \xi
+\tilde{b}+\bar{g}\bar{\beta}\right)  ^{2}$ and we end up with the simple
coordinate transformation
\begin{equation}
\xi+\tilde{b}+\bar{g}\bar{\beta}=\cos\theta\ .\label{xi2theta}%
\end{equation}
Then
\begin{align}
\tilde{h}_{1}&=1-\bar{\beta}^{2}-2\bar{m}\gamma\cot z+\frac{1}{\kappa^{2}%
}\left(  1-z\cot z\right)  \nonumber\\
&+\left(  \tilde{e}\gamma+\bar{\beta}z\right)
^{2}\left(  \cot^{2}z-1\right)  \ ,\label{h1ts}%
\end{align}
and the metric becomes
\begin{equation}
ds^{2}=-\tilde{h}_{1}dt^{2}+\frac{1}{\gamma^{2}}\left[  \frac{dz^{2}}%
{\tilde{h}_{1}}+\sin^{2}z\left(  d\theta^{2}+\sin^{2}\theta d{\varphi}%
^{2}\right)  \right]  \ .\label{stat}%
\end{equation}

The center, determined by $z=0,$ can be regular only if $\bar{m}=0$ and
$\tilde{e}=0.$

\subsection{The combined transformation}\label{combined}

In the previous sections, we have been led to transforming the Garc\'{\i}a
metric to the Wahlquist coordinates and hence to a form which is amenable for
accessing the static limit. We now combine these procedures into a direct
transformation, sidestepping the Wahlquist form, and compute the quantities
for slow rotation.

The coordinate transformations (\ref{xy2xizeta}), (\ref{zeta2r}) and
(\ref{r2z}), with $k=\gamma r_{0}$, can be combined to the single
transformation
\begin{equation}
z=ky\ .
\end{equation}

Garc\'{\i}a's parameters are expressed in terms of ours as follows,
\begin{align}
k  & =\gamma r_{0}\ \ \ \ \ ,\ \ \ \ \ \ \ m=\frac{\bar{m}}{r_{0}^{3}%
}\nonumber\\
n  & =-\frac{\tilde{b}}{r_{0}^{2}}\ \ \ ,\ \ \ \ \ \ \ \alpha=\frac{1}%
{r_{0}^{2}}\left(  \frac{1}{\kappa^{2}}-\bar{\beta}^{2}\right) \nonumber\\
e  & =\frac{\tilde{e}}{r_{0}^{2}}\ \ \ \ \ ,\ \ \ \ \ \ \ g=\frac{\bar{g}%
}{r_{0}}\label{params}\\
\beta & =\frac{\bar{\beta}}{r_{0}}\ \ \ \ \ ,\ \ \ \ \ \ \ a=\frac{C}%
{r_{0}^{2}}+\frac{1}{2r_{0}^{4}\gamma^{2}}\left(  E+\frac{1}{\kappa^{2}%
}\right) \nonumber\\
b  & =-\frac{1}{2r_{0}^{4}\gamma^{2}}\left(  E+\frac{1}{\kappa^{2}}\right)
.\nonumber
\end{align}
We now substitute the new parameters given by (\ref{params}) into the metric
form (\ref{dsgarc}) and introduce new coordinates by
\begin{equation}
y=\frac{z}{\gamma r_{0}}\ \ \ ,\ \ \ \ \ \tau=r_{0}t\ \ \ ,\ \ \ \ \ \sigma
=-r_{0}^{2}{\varphi}\ .
\end{equation}
Using Table 1, we find that the coordinate system is fixed by setting
\begin{align}
\delta & =1\ \ \ ,\ \ \ \xi_{0}=0\ ,\ \ \ \ E=1-\bar{\beta}^{2}\\
C  & =1+\bar{g}^{2}-\left(  \tilde{b}+\bar{g}\bar{\beta}\right)
^{2}\ .\nonumber
\end{align}
The regularity at the center is ensured by
\begin{equation}
\bar{m}=0\ \ \ ,\ \ \ \ \ \tilde{e}=0.
\end{equation}
The angular coordinate ${\vartheta}$ is introduced by writing
\begin{equation}
x=\cos{\vartheta}-\tilde{b}-\bar{g}\bar{\beta}\ .
\end{equation}
The coordinate $\vartheta$ only equals the $\theta$ defined in Sec. \ref{sect}
when the fluid is static. Then for small $r_{0}$ to linear order we obtain the
metric
\begin{equation}
ds^{2}=-\tilde{h}_{1}dt^{2}+\frac{dz^{2}}%
{\gamma^{2}\tilde{h}_{1}}+\frac{\sin^{2}z}{\gamma^{2}}\left[  d\vartheta^{2}+\sin^{2}\vartheta\left(
d{\varphi-\omega dt}\right)  ^{2}\right]   \label{dsinter}%
\end{equation}
with
\begin{equation}
\tilde{h}_{1}=1-\bar{\beta}^{2}+\frac{1}{\kappa^{2}}\left(  1-z\cot z\right)
+\bar{\beta}^{2}z^{2}\left(  \cot^{2}z-1\right)
\end{equation}
and
\begin{equation}
\omega=r_{0}\frac{\gamma^{2}}{\sin^{2}z}(\tilde{h}_{1}-1).
\end{equation}
The first-order calculation shows that $\omega$ does not depend on the angular
coordinate. To this order, the only component of the four-potential $A$ to
pick up a small new contribution is $A_{\varphi}$:
\begin{align}
A&=-\bar{\beta}z\cot z\,dt \nonumber\\
&+\ r_{0}\left[  \bar{\beta}\sin^{2}{\vartheta}%
({1}-z\cot z)-\bar{\beta}-\bar{g}\cos{\vartheta}\right]  d\varphi.\label{Aint}%
\end{align}
Note that to first order in $r_{0},$ the metric is independent of the magnetic
monopole charge parameter $\bar{g}.$ In fact, the monopole contribution
affects only the Maxwell equations but does not affect the gravitational equations.

\section{The electrovacuum exterior}\label{electrovacuum}%label{linv2.mws}\label{linevac2d.mws}}

The metric of the ambient electrovacuum domain, to first order in the angular
velocity, has the Reissner-Nordstrom form modified by the non-diagonal
rotation term,%

\begin{align}
ds^{2}  & =-\left(  1-2\,{\frac{m}{r}}+{\frac{{\mathrm{e}}^{2}}{{r}^{2}}%
}\right)  {\mathit{dt}}^{2}+\left(  1-2\,{\frac{m}{r}}+{\frac{{\mathrm{e}}^{2}%
}{{r}^{2}}}\right)  ^{-1}{\mathit{dr}}^{2}\nonumber\\
& +{r}^{2}\left[  \mathit{d\vartheta}^{2}+\sin^{2}\vartheta\ \left(
\mathit{d\varphi}-\,\omega{\mathit{dt}}\right)  ^{2}\right]
\end{align}
The four-potential $A=A_{\alpha}dx^{\alpha},$ where $A_{\alpha}=A_{\alpha
}\left(  r,\vartheta\right)  $, has a time component
\begin{equation}
A_{t}=-\frac{{\mathrm{e}}}{r}.\label{At}%
\end{equation}
From the $\left(  r,r\right)  $ and $\left(  r,\vartheta\right)  $ components
of Einstein's Eqs. we get no contribution to $A_{t},$ and the $\left(
t,\vartheta\right)  $ component gives $A_{r,\vartheta}=A_{\vartheta,r}$. Hence
there is a gauge in which
\begin{equation}
A_{r}=A_{\vartheta}=0.\label{Ar}%
\end{equation}
The other perturbation components are obtained in the remainder of this
section. Before doing that, we remark that for the special case of the slowly
rotating Kerr-Newman solution the forms (\ref{At}) and (\ref{Ar}) of the
four-potential components remain valid and we have
\begin{align}
A_{\varphi}^{(KN)}  & =\frac{a{\mathrm{e}}\sin^{2}\vartheta}{r}\\
\omega^{(KN)}  & =\frac{2am}{r^{3}}-\frac{a{\mathrm{e}}^{2}}{r^{4}}.\nonumber
\end{align}

We now proceed to computing the first-order contributions for a general
electrovacuum. From $\left(  \varphi,t\right)  $ component of Einstein's
equations the second-order differential equation follows:
\begin{equation}
r^{4}{\frac{d^{2}\omega}{d{r}^{2}}}+4r^{3}{\frac{d\omega}{dr}}-\frac
{4{\mathrm{e}}}{\sin^{2}\vartheta}\,{\frac{\partial A_{\varphi}}{\partial r}%
}=0.\label{Ephith}%
\end{equation}
For the uncharged case, ${\mathrm{e}}=0,$ this has the solution $\omega
=\omega_{0}+2am/r^{3}$ where $a$ is the Kerr parameter. The value of the
constant $\omega_{0}$ can be set to zero by the transformation $\varphi
\rightarrow\varphi+c_{1}t.$

The solution of (\ref{Ephith}) for a generic charge ${\mathrm{e}}\neq0$ is
\begin{equation}
A_{\varphi}=\frac{1}{4{\mathrm{e}}}\,{r}^{4}\sin^{2}\vartheta\ {\frac{d\omega
}{dr}}+f({\vartheta}).\label{Aphi1}%
\end{equation}
The Maxwell equation for the component $A_{\varphi}$ is
\begin{align}
& \left(  2\,m\ {r}-{r}^{2}-{\mathrm{e}}^{2}\right)  {\frac{\partial
^{2}A_{\varphi}}{\partial{r}^{2}}}+{r}^{2}\sin^{2}\vartheta\ {\mathrm{e}}%
\,{\frac{d\omega}{dr}}\\
& +2\,\left(  \frac{{\mathrm{e}}^{2}}{r}-m\right)  {\frac{\partial A_{\varphi}%
}{\partial r}}-{\frac{\partial^{2}{A_{\varphi}}}{\partial{\vartheta}^{2}}%
}+\cot\vartheta\,{\frac{\partial A_{\varphi}}{\partial{\vartheta}}%
}=0.\nonumber
\end{align}
Substituting $A_{\varphi}$ from (\ref{Aphi1}):
\begin{align}
& \tfrac{1}{4}\,\left(  2m\ {r}-{r}^{2}-{\mathrm{e}}^{2}\right)  {r}^{4}%
{\frac{d^{3}\omega}{d{r}^{3}}}\label{omegaeq}\\
& +\left(  \tfrac{7}{2}\,{r}m-2\,{r}^{2}-\tfrac{3}{2}{\mathrm{e}}^{2}\right)
\,{r}^{3}{\frac{d^{2}\omega}{d{r}^{2}}}\nonumber\\
& +\left(  4\,m-\tfrac{5}{2}\,{r}\right)  {r}^{3}{\frac{d\omega}{dr}%
}\nonumber\\
& ={\frac{{\mathrm{e}}}{\sin^{2}\vartheta\ }}\left(  {\frac{d^{2}f({\vartheta}%
)}{d{\vartheta}^{2}}}-\cot\vartheta\ {\frac{df({\vartheta})}{d{\vartheta}}%
}\right)  .\nonumber
\end{align}
This is a separable equation. Introducing the separation constant $K,$ we get
\begin{equation}
{\frac{{\mathrm{e}}}{\sin^{2}\vartheta\ }}\left(  {\frac{d^{2}f({\vartheta}%
)}{d{\vartheta}^{2}}}-\cot\vartheta\ {\frac{df({\vartheta})}{d{\vartheta}}%
}\right)  =K
\end{equation}
with the solution
\begin{equation}
f({\vartheta})=\frac{1}{2}{\frac{K\cos^{2}\vartheta\ }{\mathrm{e}}}%
+\frac{{\mathit{C}}_{\mathit{4}}}{\mathrm{e}}+\frac{\mathit{C}_{\mathit{5}}%
\,}{\mathrm{e}}\cos\vartheta\ .\label{f}%
\end{equation}

We next consider the radial part of Eq. (\ref{omegaeq}).

(i) For the case ${\mathrm{e}}^{2}\neq{m}^{2}$ the general solution of the
radial equation is
\begin{align}
\omega&=\mathit{C}_{\mathit{1}}+{\frac{{\mathrm{e}}^{2}-2\,mr}{3{m}^{3}{r}^{4}}%
}\mathit{C}_{\mathit{0}}+{\frac{2}{r}}\mathit{C}_{\mathit{2}} \nonumber\\
&+{\frac
{\,{\mathrm{e}}^{2}\left(  {r}^{2}-2\,mr+{\mathrm{e}}^{2}\right)  \left[
r+m+\left(  {\mathrm{e}}^{2}-{r}^{2}\right)  L\right]  }{{m}^{2}\left(
{\mathrm{e}}^{2}-{m}^{2}\right)  ^{2}{r}^{4}}}\mathit{C}_{\mathit{3}%
}\label{omega}%
\end{align}
where
\begin{equation}
C_{0}=K{m}^{2}+3\,{\mathrm{e}}^{2}{m}^{2}\mathit{C}_{\mathit{2}}+\frac
{2\,{\mathrm{e}}^{2}}{{\mathrm{e}}^{2}-{m}^{2}}\mathit{C}_{\mathit{3}}%
\end{equation}
and
\begin{equation}
L\left(  r\right)  =\left\{  {\
\begin{tabular}
[c]{l}%
$\frac{1}{2\sqrt{{m}^{2}-{\mathrm{e}}^{2}}}\mathit{\ln}{\frac{r-m+\sqrt{{m}%
^{2}-{\mathrm{e}}^{2}}}{r-m-\sqrt{{m}^{2}-{\mathrm{e}}^{2}}}\;\;\;\;\;\qquad
\qquad}$if $m>{\mathrm{e}}$\\
$\frac{1}{\sqrt{{\mathrm{e}}^{2}-{m}^{2}}}\left(  \frac{\pi}{2}-\mathit{\arctan
}{\frac{r-m}{\sqrt{{\mathrm{e}}^{2}-{m}^{2}}}}\right)  {\,\qquad\;\;}$if
$m<{\mathrm{e}}$%
\end{tabular}
\ \ }\right. \label{L}%
\end{equation}
with the derivative
\begin{equation}
\frac{dL\left(  r\right)  }{dr}=\frac{1}{2\,m\ {r}-{r}^{2}-{\mathrm{e}}^{2}%
}.\label{dL}%
\end{equation}

\bigskip(ii) For the equilibrium case ${\mathrm{e}}^{2}={m}^{2},$ the radial
solution is
\begin{align}
\omega&=\mathit{C}_{\mathit{1}}+{\frac{m-2\,r}{3{r}^{4}}}\left(  K+3\,{m}%
^{2}\mathit{C}_{\mathit{2}}\right)  +{\frac{2}{r}}\mathit{C}_{\mathit{2}%
} \nonumber\\
&+\frac{2}{15}\frac{5r^{2}-4mr+m^{2}}{m(r-m)^{2}r^{4}}\mathit{C}_{\mathit{3}%
}.\label{omeq}%
\end{align}
When an asymptotically nonrotating frame is chosen, we have that
$\mathit{C}_{\mathit{1}}=0,$ and we shall assume this to hold in the sequel.

Comparing with the Kerr-Newman metric with rotation parameter $a,$ we obtain
\[
K=-3am.
\]

Substituting the solution (\ref{omega}), (\ref{f}) and (\ref{dL}) in the
potential (\ref{Aphi1}) we get for the case ${{\mathrm{e}}^{2}}\neq{m}^{2}:$%
\begin{align}
A_{\varphi}  & =\frac{a{{\mathrm{e}}}}{r}\sin^{2}\vartheta+\frac{3{\mathrm{e}}%
^{2}mr-mr^{3}-2{{\mathrm{e}}^{4}}}{2{{\mathrm{e}}}mr}C_{2}\sin^{2}\vartheta\nonumber\\
& +\frac{3m^{2}r(m+r)-2{{\mathrm{e}}^{2}}m^{2}-4{{\mathrm{e}}^{4}}}{6m^{3}r\left(
{{\mathrm{e}}^{2}}-{m}^{2}\right)  ^{2}}C_{3}{{\mathrm{e}}}\sin^{2}\vartheta
\label{Aphi2}\\
& +\frac{3{{\mathrm{e}}^{2}}mr-mr^{3}-2{{\mathrm{e}}^{4}}}{2m^{2}r\left(
{{\mathrm{e}}^{2}}-{m}^{2}\right)  ^{2}}C_{3}{\mathrm{e}}L\left(  r\right)  \sin
^{2}\vartheta \nonumber\\
&-\frac{3am}{2{\mathrm{e}}}\,{+}\frac{{C}_{4}}{\mathrm{e}}%
+\frac{C_{5}}{\mathrm{e}}\cos\vartheta.\nonumber
\end{align}
For ${\mathrm{e}}=\pm{m}$ , we get identical limits from both values of $L$ in
(\ref{L})\textbf{. }The limiting form of the potential $A_{\varphi}$ can be
obtained by substituting (\ref{omeq}) in Eq. (\ref{Aphi1}):
\begin{align}
\pm A_{\varphi}  & =\frac{am}{r}\sin^{2}\vartheta+\frac{\left(  r+2m\right)
\left(  r-m\right)  ^{2}}{2mr}C_{2}\sin^{2}\vartheta\\
& -\frac{\left(  2r-m\right)  \left(  2m^{2}-5mr+5r^{2}\right)  }%
{15m^{2}r\left(  r-m\right)  ^{3}}C_{3}\sin^{2}\vartheta \nonumber\\
&-\frac{3a}{2}%
\,{+}\frac{{C}_{4}}{m}+\frac{C_{5}}{m}\cos\vartheta.\nonumber
\end{align}
The constant $C_{4}$ is inessential since it does not appear in the Maxwell tensor.

\bigskip In order to clarify the r\^{o}le of the constants $C_{2}$ and $C_{5}
$, it will be helpful to introduce a tetrad with the components%
\begin{align}
e_{0}  & =\left(  1-2\,{\frac{m}{r}}+{\frac{{\mathrm{e}}^{2}}{{r}^{2}}}\right)
^{-1/2}\frac{\partial}{\partial t}\\
e_{1}  & =\left(  1-2\,{\frac{m}{r}}+{\frac{{\mathrm{e}}^{2}}{{r}^{2}}}\right)
^{1/2}\frac{\partial}{\partial r}\nonumber\\
e_{2}  & =\frac{1}{r}\frac{\partial}{\partial\vartheta}\nonumber\\
e_{3}  & =\frac{1}{r\sin\vartheta}\frac{\partial}{\partial\varphi}-\omega
r\sin\vartheta\left(  1-2\,{\frac{m}{r}}+{\frac{{\mathrm{e}}^{2}}{{r}^{2}}%
}\right)  ^{-1}\frac{\partial}{\partial t}.\nonumber
\end{align}
In this tetrad the leading terms for large $r$ in the components of the
Riemann tensor are given by
\begin{align}
R_{0113}  & =R_{0223}=\frac{C_{2}\sin\vartheta}{r^{2}}\\
\frac{1}{2}R_{0123}  & =R_{0213}=-R_{0312}=\frac{C_{2}\cos\vartheta}{r^{2}%
}\nonumber
\end{align}
As seen the terms do not fall off sufficiently fast for the space-time to be
asymptotically flat. Thus this object does not seem to be isolated. To see
this we look at the asymptotic behaviour of the electromagnetic field in this
tetrad. \bigskip From Eq. (\ref{Aphi2}) we obtain
\begin{align}
\lim_{r\rightarrow\infty}F_{13}  & =-\frac{C_{2}}{\mathrm{e}}\sin
\vartheta\equiv-B_{2}\\
\lim_{r\rightarrow\infty}F_{23}  & =-\frac{C_{2}}{\mathrm{e}}\cos
\vartheta\equiv B_{1}.\nonumber
\end{align}
This can be interpreted that for nonzero values of $C_{2}$, the fluid ball is
immersed in a constant (in this approximation) external magnetic field,
parallel to the axis of rotation, and extending to infinity. Note that the
Riemann tensor falls off even though the electromagnetic field tends to a
constant value. One would expect that the contribution of the electromagnetic
stresses to the curvature will appear in a higher approximation. \bigskip

Hence we have the following

\textbf{Theorem.}

\textit{When }$C_{2}$\textit{\ has a nonzero value, the external solution is
not asymptotically flat. }

It is therefore important to investigate the values of $C_{2}$ that the
matching provides. This investigation will be carried out in the next section.

The term containing $C_{5}$ is the potential of a magnetic monopole. In the
given frame, it gives rise to a purely radial magnetic field which has the
asymptotic form $\thicksim1/r^{2}$.

\section{Matching}\label{match}

In this section, the matching procedure is described. The next subsection
outlines the method in general terms. We then join the static, spherically
symmetric external and internal domains in subsection \ref{sphere} The static internal
state is parameterized by the constants $\bar{\beta},$ $\kappa$ and $\gamma$.
As a result of the matching, these three constants determine the radius
$r_{1}$ of the matching surface and the parameters $m$ and $\mathrm{e}$ of the
vacuum exterior. It is a consequence of the matching conditions that the
surface of matching coincides with the zero pressure surface of the interior.
In subsection \ref{first} we carry out the matching to first order in the rotation
parameter. This will yield the parameters of the slowly rotating electrovacuum
region in terms of $r_{0},$ the parameter describing the angular velocity of
the fluid, and in terms of the parameter $\bar{g}.$

\subsection{The matching conditions}\label{cond}

We want a global model from which (i) any surface charges and currents,
furthermore (ii) surface layers of matter are absent. From the first condition
it follows\cite{MTW} that the electromagnetic stress tensor can be
continuously matched at the surface $\Sigma$ if we assume that both the
permeablity and dielectric coefficients are equal in the electrovacuum and in
the interior. This means with (ii) that there is no discontinuity in the
pressure $p$ across $\Sigma$.

We write the metric $ds^{2}\equiv g_{\alpha\beta}dx^{\alpha}dx^{\beta}$ for
both the interior and the exterior regions in curvature coordinates, $\left\{
x^{\alpha}\right\}  =\left\{  t,r,\vartheta,\varphi\right\}  ,$ in the
following form:
\begin{equation}
ds^{2}=-{\mathcal{A}}^{2}dt^{2}+{\mathcal{B}}^{2}dr^{2}+r^{2}\left[
d\vartheta^{2}+\sin^{2}\vartheta\left(  d\varphi-\omega dt\right)
^{2}\right] \label{dsuniv}%
\end{equation}
where $\mathcal{A}$, $\mathcal{B}$ and ${\omega}$ are functions of the radial
coordinate $r$ alone. Both to order zero and one in the angular velocity, the
constant-pressure surfaces of the perfect fluid coincide with the constant $r$ surfaces.

We match the hypersurface given by $r=r_{1}$ , of the interior region, with
the corresponding matching surface at $r=r_{1}$ , of the exterior region, such
that the induced metrics $ds^{2}|_{\Sigma}$ and induced extrinsic curvatures
$K|_{\Sigma}$ are equal. The continuity of the functions $\mathcal{A}$ and
$\omega$ across the matching surface can be achieved by transforming the
coordinates such that
\begin{equation}
t=C_{6}t^{\prime},\qquad\varphi=\varphi^{\prime}+\Omega t^{\prime}%
\end{equation}
where $C_{6}$ and $\Omega$ are suitably chosen constants and $x^{1}=r$ and
$x^{2}=\vartheta$ are unchanged. We shall, however drop the primes from the
new coordinates.

The normal of the hypersurface $\Sigma$ has the form
\begin{equation}
n=\frac{1}{\mathcal{B}}\frac{\partial}{\partial r}%
\end{equation}
The extrinsic curvature has the nonvanishing components:\cite{bfmp}
\begin{align}
K_{00}  & \equiv\tfrac{1}{2}g_{00,1}n^{1}=-\frac{\mathcal{A}}{\mathcal{B}%
}{\mathcal{A}}_{,r}\label{K00}\\
K_{03}  & \equiv\tfrac{1}{2}g_{03,1}n^{1}=-\frac{1}{2\mathcal{B}}\sin
^{2}\vartheta\left(  r^{2}\omega\right)  _{,r}\label{K03}\\
K_{22}  & \equiv\tfrac{1}{2}g_{22,1}n^{1}=\frac{r}{\mathcal{B}}\label{K22}\\
K_{33}  & \equiv\tfrac{1}{2}g_{33,1}n^{1}=\frac{r}{\mathcal{B}}\sin
^{2}\vartheta.\label{K33}%
\end{align}
The junction of $K_{22}$ with Eq. (\ref{K22}) implies that the function
$\mathcal{B}$ must be ${\mathcal{C}}^{0}$ at $r=r_{1}.$ (This implies the
matching of $K_{33}$ .) Next we conclude from Eqs. (\ref{K00}) and
(\ref{K03}), respectively, that $\mathcal{A}$ and $\omega$ are ${\mathcal{C}}%
^{1}$ functions at $r_{1}.$

The four-potential $A=A_{\alpha}dx^{\alpha}$ in both regions has a small
$A_{\varphi}$ component, and $A_{r}=A_{\vartheta}=0.$ The timelike component
$A_{t}$ retains its spherically symmetric form to first order in the angular
velocity. From condition (i), the following components of the electromagnetic
stress tensor $F_{\alpha\beta}$ are continuous across the matching surface
$\Sigma:$ $F_{\alpha\beta}h_{\gamma}^{\alpha}h_{\delta}^{\beta}\ $and
$F_{\alpha\beta}h_{\gamma}^{\alpha}n^{\beta},$ where $h_{\gamma}^{\alpha
}=\delta_{\gamma}^{\alpha}-n^{\alpha}n_{\gamma}$ is the projection tensor into
the orthogonal complement to the normal $n$ of the tangent space\footnote{If
we allow for surface currents and charges (but still assume that the
permeability and dielectricity are that of the vacuum) these conditions take
the general form $\left[  F_{\alpha\beta}\right]  h_{\gamma}^{\alpha}%
h_{\delta}^{\beta}=0$ and $\left[  F_{\alpha\beta}\right]  h_{\gamma}^{\alpha
}n^{\beta}=4\pi j_{\gamma}^{\left(  surface\right)  }$ where the jump is
indicated by a bracket.}. We must match the following nonvanishing
components:
\begin{align}
F_{rt}  & =A_{t,r}\label{Atr}\\
F_{r\varphi}  & =A_{\varphi,r}\label{Apr}\\
F_{\vartheta\varphi}  & =A_{\varphi,\vartheta}.\label{Aph}%
\end{align}

\subsection{Spherical matching}\label{sphere}%\label{maxmat4.mws}}

We first carry out the matching of the electrovacuum region at the sphere
$r=r_{1}$ with the internal region at $z=z_{1}$ to order zero in the rotation
parameter $r_{0}$. The metric of the perfect fluid takes a simpler form when
using the coordinate $z.$ However, the matching process is more transparent
when using the radial coordinate $r.$ These dual pictures are connected by the
transformation $\gamma r=\sin z$ [\textit{Cf.} (\ref{r2z})]. From the
continuity of the metric component $g_{\vartheta\vartheta}$ at the junction
surface we find
\begin{equation}
r_{1}=\frac{\sin z_{1}}{\gamma}.
\end{equation}
Eliminating the mass $m$ between the junction conditions of the metric
components $g_{tt}$ and $K_{\vartheta\vartheta},$ we get the simple result for
the value of the parameter%

\begin{equation}
C_{6}=\cos z_{1}.
\end{equation}
Continuity of the radial component of the electromagnetic field yields the
value of the total charge,
\begin{equation}
{\mathrm{e}}=\tfrac{\bar{\beta}}{\gamma}(z_{1}-\sin z_{1}\cos z_{1}).\label{eeq}%
\end{equation}
We next eliminate $m$ between the junction condition of $g_{tt}$ and that of
the extrinsic curvature component $K_{tt}.$ Solving for $\kappa^{2}$ , we get
\begin{equation}
\frac{1}{\kappa^{2}}=\frac{\tan z_{1}}{z_{1}}+2\bar{\beta}^{2}\left(  2+z_{1}%
\cot2z_{1}\right)  .\label{kaeq}%
\end{equation}
The mass is then obtained from the condition of continuity of the $g_{tt}$
component as%

\begin{align}
m&=\frac{r_{1}}{2}\left(  1-\frac{\cos^{2}z_{1}}{\kappa^{2}}\right) \nonumber\\
&+\frac{\bar{\beta}^{2}}{2\gamma\sin z_{1}}\left(  z_{1}^{2}+z_{1}\sin2z_{1}%
\cos2z_{1}+\tfrac{1}{2}\sin^{2}2z_{1}\right)  .\label{meq}%
\end{align}

The specific charge of the body is characterized by the function
$m^{2}-{\mathrm{e}}^{2},$ given by Eqs. (\ref{eeq}), (\ref{kaeq}) and
(\ref{meq}). Taylor expanding about the origin, $z_{1}=0,$ we get
\begin{equation}
m^{2}-{\mathrm{e}}^{2}=\frac{1-4\bar{\beta}^{2}}{9\gamma^{2}}z_{1}^{6}+{\mathcal{O}}%
\left(  z_{1}^{8}\right)  .
\end{equation}
Hence we find that for small stars, ${\mathrm{e}}^{2}>m^{2}$ for the values
$\left\vert \bar{\beta}\right\vert >1/2.$ In other cases, we need to treat the
different types of specific charge separately.

\begin{figure}
%\hskip 3cm
\epsfxsize=3in
\epsffile{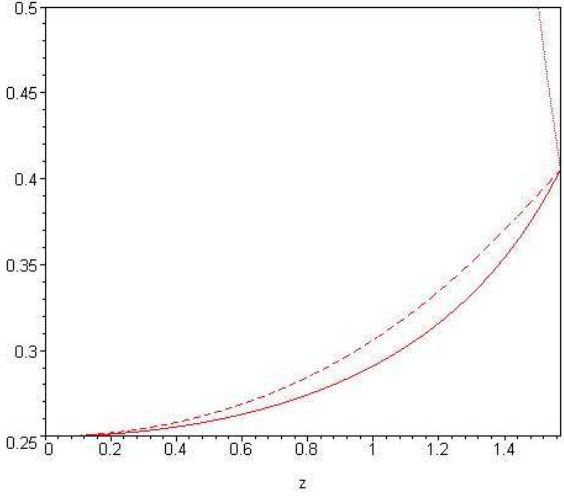} 
\caption{The physically allowed region of the specific charge $\bar{\beta}^2$: The solid curve gives the maximally allowed value as a function of the radius.}
\label{figure:garciabeta}
\end{figure}

The pressure at the center is required to be non-negative. On Fig. \ref{figure:garciabeta}, we plot
$\bar{\beta}^{2}$ as a function of the radius $z_{1}$ of the star for the limiting
case when the pressure at the center vanishes (solid curve). The allowed
region of $\bar{\beta}^{2}$ lies under this curve. The other two curves represent
the two solutions for $\bar{\beta}^{2}$ for the extremely charged star,
$m^{2}={\mathrm{e}}^{2}.$ However, the extremely charged state lies outside the
physically allowed domain.%

\subsection{First-order matching}\label{first}

From the continuity of the metric component $g_{t\varphi}$ we get the angular
velocity of the fluid:
\begin{align}
\Omega & =\tfrac{r_{0}{\gamma}^{2}}{\sin^{2}z_{1}}\left[  {\bar{\beta}}^{2}z_{1}^{2}(\cot
^{2}z_{1}-1)-{\bar{\beta}}^{2}+{\kappa}^{-2}(1-z_{1}\cot z_{1})\right]
\nonumber\\
&+ \tfrac{{\gamma}^{3}(2m\sin z_{1}-\gamma{\mathrm{e}}^{2})}{m^{3}\sin^{4}z_{1}\cos z_{1}}%
({\mathrm{e}}^{2}m^{2}C_{2}+\tfrac
{2{\mathrm{e}}^{2}}{3\left({\mathrm{e}}^{2}-m^{2}\right)}C_{3}-am^{3})\nonumber\\
& -C_{3}\tfrac{{\gamma}^{4}}{m^{2}\sin^{4}z_{1}\cos z_{1}}\tfrac
{{\mathrm{e}}^{2}}{({\mathrm{e}}^{2}-m^{2})^{2}}({\mathrm{e}}^{2}+\tfrac{\sin
^{2}z_{1}}{{\gamma}^{2}}-2m\tfrac{\sin z_{1}}{{\gamma}})\nonumber\\
& \times\left[  \tfrac{\sin z_{1}}{{\gamma}}+m+({\mathrm{e}}^{2}-\tfrac{\sin
^{2}z_{1}}{{\gamma}^{2}})L(\tfrac{\sin z_{1}}{{\gamma}})\right]-\tfrac{4{\gamma}}{\sin2z_{1}}C_{2}
\end{align}

\begin{figure}
\vskip 0.7cm
\epsfxsize=3in
\epsffile{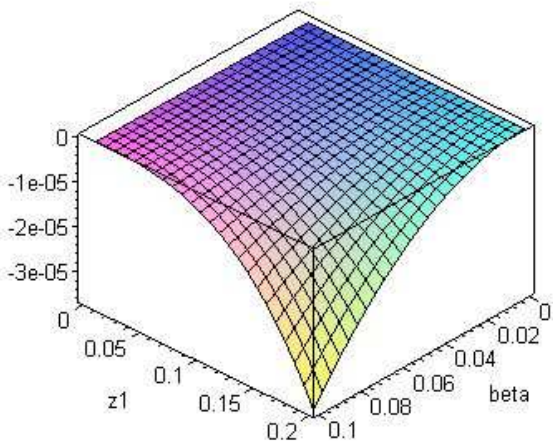} 
\caption{The coefficient $C_2$ as a function of the specific charge $\bar{\beta}$ and the radius $z_1$. }
\label{figure:garciafig}
\end{figure}

The condition of continuity of the Maxwell field $F_{\varphi\vartheta}$ can be
investigated by using the form of the four-potential in Eqs. (\ref{Aint}) and
(\ref{Aphi2}). The matching of the magnetic monopole terms, proportional to
$\cos\vartheta,$ yields that
\begin{equation}
C_{5}=-r_{0}{\mathrm{e}}\bar{g}.
\end{equation}
The rest of this matching equation, taken together with the continuity of
$F_{\varphi r}$ and $K_{t\varphi}$ can be solved for the parameters $a,$
$C_{2}$ and $C_{3}$ as follows,
\begin{align}
a  & =\frac{r_{0}}{6{\gamma} m {\kappa}^{2}}\left\{  {\mathrm{e}}%
\bar{\beta}^{-1}{\gamma}\left(  4\bar{\beta}^{2}{\kappa}^{2}-3\right)
-2{\bar{\beta}}^{2}{\kappa}^{2}z_{1}\cos2z_{1}\right. \\
& \left.  +2z_{1}\sin^{2}z_{1}\,+\bar{\beta}\,{\kappa}^{2}\,4\,z_{1}\,\left(
z_{1}\,\bar{\beta}-{\mathrm{e}}\,{\gamma}\right)  \,\cot z_{1} \right. \nonumber\\
& \left. -\left(  1+2\,z_{1}%
^{2}\right)  \,\bar{\beta}\,\sin2z_{1}\right\}  .\nonumber
\end{align}%
\begin{align}
C_{2}  & =C_{L0}-16r_{0}{\mathrm{e}}\gamma\bar{\beta}^{3}z_{1}^{2}\cos^{2}z_{1}\sin^{4}%
z_{1}\frac{FGH}{D^{2}}L\left(  \frac{\sin z_{1}}{\gamma}\right)  \\
C_{3}  & =\frac{r_{0}\,{\bar{\beta}}^{4}\,\cos z_{1}\,\cot z_{1}\,}{4096\,{\gamma
}^{4}\,z_{1}^{3}\,{\mathrm{e}}}FG\,{H}^{2}.
\end{align}
The detailed form of the quantities here is
\begin{align}
C_{L0}  & =\,\,\frac{{\bar{\beta}}{\mathrm{e}}{\gamma}^{2}r_{0}}{3D^{2}}\,z_{1}\tan
z_{1}\{-512{\mathrm{e}}^{4}\bar{\beta}^{-4}\gamma^{4}{\sin}^{6}{z_{1}}\nonumber\\
&-70\,\bar{\beta}
^{2}z_{1}\,{\sin{^{4}}z_{1}}\sin2z_{1} \nonumber\\
& +4\bar{\beta}^{2}z_{1}^{2}\,{\sin{^{4}}z_{1}}\left(  39+448\,z_{1}^{2}%
+384\,z_{1}^{4}\right) \nonumber\\
& +7\bar{\beta}^{2}z_{1}\,{\sin{^{4}}z_{1}}\left(  5\,\sin6z_{1}-\sin10\,z_{1}%
\right) \nonumber\\
& +2\bar{\beta}^{2}z_{1}^{2}\,{\sin{^{4}}z_{1}}\left(  \,6\cos2z_{1}+144\,z_{1}%
^{2}\cos2z_{1}\right) \nonumber\\
& -2\bar{\beta}^{2}z_{1}^{2}\,{\sin{^{4}}z_{1}}\left(104\cos4z_{1}+  \,9\,\cos6z_{1}%
\right) \nonumber\\
& +2\bar{\beta}^{2}z_{1}^{2}\,{\sin{^{4}}z_{1}}\left(
26\,\cos8\,z_{1}+3\,\cos10\,z_{1}\right) \nonumber\\
& -64\,\bar{\beta}^{2}z_{1}^{3}\,{\sin{^{4}}z_{1}}\,\left(  28z_{1}\,\cos
4z_{1}+21{\sin}^{3}{2z_{1}}\right) \nonumber\\
& -32\,\bar{\beta}^{2}z_{1}^{3}\,{\sin{^{4}}z_{1}}\,\left(12\cos2z_{1}\,{\sin}^{3}{2z_{1}}+ 9\,z_{1}\cos
6z_{1}\right) \nonumber\\
& -32\,\bar{\beta}^{2}z_{1}^{3}\,{\sin{^{4}}z_{1}}\,\left(
128\,z_{1}^{2}\,\sin2z_{1}+24\,z_{1}^{2}\,\sin4z_{1}\right) \nonumber\\
& -\bar{\beta}^{4}z_{1}^{2}\,{\sin^{2}z_{1}}\left(  40-162\,z_{1}^{2}%
+1136\,z_{1}^{4}+1536\,z_{1}^{6}\right) \nonumber\\
& -8\bar{\beta}^{4}z_{1}^{2}\,{\sin^{2}z_{1}}\,\left(  z_{1}^{2}+56\,z_{1}%
^{4}\right)  \,\cos2z_{1}\nonumber\\
& -\bar{\beta}^{4}z_{1}^{2}\,{\sin^{2}z_{1}}\cos4z_{1}\left(  221\,z_{1}%
^{2}-1088\,z_{1}^{4}-60\right)  \,\nonumber\\
& -4\,\bar{\beta}^{4}z_{1}^{2}\,{\sin^{2}z_{1}}\left(  6\,\cos8\,z_{1}%
-\cos12\,z_{1}\right) \nonumber\\
& +32\bar{\beta}^{4}z_{1}^{3}\,{\sin^{2}z_{1}}\,\left(  7\,\cos2z_{1}-2\right)
\,{\sin}^{5}{2z_{1}}\,\\
& +4\bar{\beta}^{4}z_{1}^{4}\,{\sin^{2}z_{1}} \left(  3+112\,z_{1}%
^{2}\right)  \cos6z_{1} 
\nonumber\\
& - 3\bar{\beta}^{4}z_{1}^{4}\,{\sin^{2}z_{1}}\cos12\,z_{1}
\nonumber\\
& +256\,\bar{\beta}^{4}z_{1}^{7}\,{\sin^{2}z_{1}}\,\left(  10\,\sin2z_{1}%
+\sin4z_{1}\right) \nonumber\\
& +128\,\,\bar{\beta}^{4}z_{1}^{5}\,{\sin^{2}z_{1}}\left(  5+6\,\cos2z_{1}\right)
\,{\sin}^{3}{2z_{1}}\nonumber\\
& +2\bar{\beta}^{4}z_{1}^{4}\,{\sin^{2}z_{1}}\left[  \,\left(  31+24\,z_{1}%
^{2}\right)  \cos8\,z_{1}-2\,\cos10\,z_{1}\right] \nonumber\\
& +\bar{\beta}^{6}z_{1}^{3}\,{B}^{2}\,\left(  32\,z_{1}^{3}-12\,z_{1}-9\,\sin
2z_{1}+3\,\sin6z_{1}\right) \nonumber\\
& +2\,\bar{\beta}^{6}z_{1}^{4}\,{B}^{2}\,\,\left(  \cos2z_{1}+6\,\cos4z_{1}%
-\cos6z_{1}\right)  \nonumber\\
& +128\,\bar{\beta}^{6}z_{1}^{5}\,{B}^{2}\,\,{\cos}^{3}{z_{1}}\sin z_{1}  \}\nonumber
\end{align}
where
\begin{align*}
D  & =16{\mathrm{e}}^{2}\,\bar{\beta}^{-2}{\gamma}^{2}{\sin^{4}z_{1}}+\bar{\beta}
^{4\,}z_{1}^{2}\,{B}^{2}\\
& +2\bar{\beta}^{2}z_{1}\,{\sin^{2}z_{1}}\left[  \,\left(  3+\,24z_{1}^{2}\right)
\sin2z_{1}-\sin6z_{1}\right] \\
& +2\bar{\beta}^{2}z_{1}^{2}\,{\sin^{2}z_{1}}\left(  4z_{1}\sin4z_{1}-\cos
2z_{1}+8\cos4z_{1}\right) \\
& +2\bar{\beta}^{2}z_{1}^{2}\,{\sin^{2}z_{1}}\left( \cos6z_{1}-8\,-16\,z_{1}^{2}\right)
\end{align*}
and
\begin{align*}
B  & =4\,z_{1}^{2}+\cos4z_{1}+z_{1}\sin4z_{1}-1\\
F  & =4{\bar{\beta}}^{2}\,z_{1}^{2}\,+\left(  1+2\,{\bar{\beta}}^{2}z_{1}^{2}\,\right)
\,\cos2z_{1}-3\,{\bar{\beta}}^{2}z_{1}\,\,\sin2z_{1}-1\\
G  & =\left(  8\,z_{1}^{2}-1\right)  \,\cos z_{1}+\cos3z_{1}-4\,z\,_{1}\sin
z_{1}\\
H  & =2\,z_{1}\,\left(  2+\,{\bar{\beta}}^{2}\,B\right)  -4\,z_{1}\,\cos
2z_{1}-2\,\sin2z_{1}+\sin4z_{1}.
\end{align*}
We next investigate the value of the constant $C_{2}$ as a function of the
radius $z_{1}$ and the charge density parameter $\,{\bar{\beta}.}$ 
The values of $C_{2}$ are strictly negative in the physically allowed region
$\left\{  z_{1}\in\left(  0,\tfrac{\pi}{2}\right)  ,{\bar{\beta}\in}\left(
0,0.5\right)  \right\}  .$ 
On Fig. \ref{figure:garciafig} we
display the values of $C_{2}$ on a 3-dimensional diagram. For clarity we have chosen
the parameter
region $\left\{  z_{1}\in\left(  0,0.2\right)  ,{\bar{\beta}\in}\left(
0,0.1\right)  \right\}  .$ (The values have been obtained by the software
\textit{Maple} of the University of Waterloo, using a precision of $33$
significant digits).

Based on this numerical result, we are in position to state the following

\textbf{Conjecture}:

\textit{The Garc\'{\i}a solution cannot be used as the model of an isolated
rotating body}.

Our conjecture does not prohibit physical applications of the Garcia solution.
For a nonvanishing parameter $C_{2},$ the fluid domain is embedded in an
external homogeneous magnetic field parallel to the axis of rotation. This is
a typical setting in the interior of galactic disks.

$\smallskip$

\section{Discussion of results}\label{discussion}

The Garc\'{\i}a solution is the electrically charged generalization of the
Wahlquist space time and it carries the extra parameter $\bar{\beta}$ determining
the charge density. The new degrees of freedom that the presence of electric
charge bring in the model would seem to raise the possibility of a successful
matching to an empty exterior domain. However, the appearance of these new
degrees of freedom is compensated by a larger number of matching conditions.
In addition to the surface gravity and first curvature, one must match also
the electromagnetic field, to rule out surface charges and currents. As we
demonstrated in the main part of this paper, the net effect is that the
charged generalization of the Wahlquist metric is even less likely to serve as
a model of an isolated star. An alternative approach would be to assume that
the dielectricity or the permeability of the interior solution differs from
that of the vacuum. We could then solve the junction conditions for the
electric polarizability $\vec{P}$ and/or magnetization $\vec{M}$ . However
this would probably not give rise to any physically realistic model of the
substance occupying the interior region.

\acknowledgments

This research has been supported by the {\it OTKA} grant T031724. MB has been
supported by a grant from the {\it International Centre for Workshops in
Theoretical Physics} and by the {\it Swedish Research Council}.

\section{Appendix}\label{appendix}

An important question is how many physical parameters there are in the metric
(\ref{dsgarc}). A coordinate transformation $\tau=\tau^{\prime}+c\sigma,$
where $c$ is a constant, can be used to set $\xi_{0}$ to an arbitrary value. A
transformation $\sigma=c\sigma^{\prime}$ rescales the constant $\delta.$ The
determinant of the $(\tau,\sigma)$ part of the metric is $-\delta^{2}PQ,$
showing that the symmetry axis is at those values of $x=x_{0}$ where
$P(x_{0})=0$. The coefficient of the $d\tau d\sigma$ part of the metric is
$\frac{\delta}{\Delta}(PN-QM)$. In order to obtain a coordinate system which
can be made regular at the axis we must have $M(x_{0})=0$, i.\ e.
\[
\xi_{0}=\frac{1}{k}\sinh(kx_{0})\ .
\]
The constant $\delta$ should be set by requiring the usual perimeter per
radius ratio $2\pi$ for small circles near the axis, taking into account the
range of the cyclical coordinate $\sigma$.

We should decide whether or not two different sets of constants $a,$ $b,$ $e,
$ $g,$ $k,$ $m,$ $n,$ $\alpha,$ $\beta,$ $\delta$ and $\xi_{0}$ determine
different spacetimes. Let us assume for the time being that we set $\xi_{0}=0$
and $\delta=1$. If we perform the coordinate transformation $x^{\prime}=cx,$
$y^{\prime}=cy$ and introduce the constant $k^{\prime}=k/c $ then the
functions $M$ and $N$ transform as
\[
M=\frac{1}{k^{2}}\sinh^{2}(kx)=\frac{1}{c^{2}k^{\prime2}}\sinh^{2}(k^{\prime
}x^{\prime})=\frac{1}{c^{2}}M^{\prime}%
\]
and $N=N^{\prime}/c^{2}.$ Introducing the constants $a^{\prime}=c^{4}a,$
$b^{\prime}=c^{4}b,$ $e^{\prime}=c^{2}e,$ $g^{\prime}=c^{2}g,$ $m^{\prime
}=c^{3}m,$ $n^{\prime}=c^{3}n,$ $\alpha^{\prime}=c^{2}\alpha,$ $\beta^{\prime
}=c\beta,$ and defining $P^{\prime}$ and $Q^{\prime}$ similarly to (\ref{fcs})
for $P$ and $Q,$ we get
\[
P=\frac{1}{c^{4}}P^{\prime}\ \ ,\ \ \ Q=\frac{1}{c^{4}}Q^{\prime}\ .
\]
Then the $\left(  x,y\right)  $ bloc of the metric becomes
\[
\Delta\left(  \frac{dx^{2}}{P}+\frac{dy^{2}}{Q}\right)  =\Delta^{\prime
}\left(  \frac{dx^{\prime2}}{P^{\prime}}+\frac{dy^{\prime2}}{Q^{\prime}%
}\right)  \ .
\]
To get the appropriate transformation for the other terms in the metric, we
have to change the coordinates as $\tau=c\tau^{\prime}$ and $\sigma
=c^{3}\sigma^{\prime}.$ It follows that the two sets of constants, $a,$ $b, $
$e,$ $g,$ $k,$ $m,$ $n,$ $\alpha,$ $\beta$ and $a^{\prime},$ $b^{\prime},$
$e^{\prime},$ $g^{\prime},$ $k^{\prime},$ $m^{\prime},$ $n^{\prime}, $
$\alpha^{\prime},$ $\beta^{\prime}$ describe diffeomorphic metrics. Since no
other similar freedom exists, there are eight physical parameters in the
charged Wahlquist metric. One can use the above freedom to set one of the
constants to some prescribed value, for example one may set $k=1$ or
$\beta=1.\;$The constants $a$ and $b$ can only be scaled by a positive
constant. Hence they may be put equal to $\pm1$ or $0$.

\end{document}